\documentclass[runningheads]{llncs}
\usepackage[dvipsnames]{xcolor}
\usepackage[T1]{fontenc}
\usepackage{subcaption}
\newcommand\ModelName{SurgWM}

\usepackage{graphicx,verbatim}
\usepackage{amsmath}
\usepackage{booktabs}
\usepackage{multirow}
\usepackage{multicol}
\usepackage{float}
\usepackage{url}
\usepackage{hyperref}

\begin{document}
\title{Surgical Vision World Model}
%
% \begin{comment}  
%% Removed for anonymized MICCAI 2025 submission
\author{Saurabh Koju\inst{3} \and
Saurav Bastola\inst{3}  \and
Prashant Shrestha\inst{3}  \and
Sanskar Amgain\inst{3}  \and
Yash Raj Shrestha\inst{4}  \and
Rudra P.K. Poudel \inst{2}  \and
Binod Bhattarai\inst{1} }
\authorrunning{Koju and Bastola et al.}
% First names are abbreviated in the running head.
% If there are more than two authors, 'et al.' is used.
%
\institute{University of Aberdeen, UK \\
\email{binod.bhattarai@abdn.ac.uk}
\and
Cambridge Research Laboratory, Toshiba Europe Ltd, UK
\and 
Nepal Applied Mathematics and Informatics Institute for research (Naamii), Nepal 
\and 
University of Lausanne, Switzerland
% \email{lncs@springer.com}
% \\
% \url{http://www.springer.com/gp/computer-science/lncs} 
% \email{\{abc,lncs\}@uni-heidelberg.de}
}

% \end{comment}

% \author{Anonymized Authors}  %% Added for anonymized MICCAI 2025 submission
% \authorrunning{Anonymized Author et al.}
% \institute{Anonymized Affiliations \\
%     \email{email@anonymized.com}}

\maketitle              % typeset the header of the contribution
\begin{abstract}
Realistic and interactive surgical simulation has the potential to facilitate crucial applications, such as medical professional training and autonomous surgical agent training. In the natural visual domain, world models have enabled action-controlled data generation, demonstrating the potential to train autonomous agents in interactive simulated environments when large-scale real data acquisition is infeasible. However, such works in the surgical domain have been limited to simplified computer simulations, and lack realism. Furthermore, existing literature in world models has predominantly dealt with action-labeled data, limiting their applicability to real-world surgical data, where obtaining action annotation is prohibitively expensive. Inspired by the recent success of Genie in leveraging \textit{unlabeled} video game data to infer latent actions and enable action-controlled data generation, we propose the first surgical vision world model. The proposed model can generate action-controllable surgical data and the architecture design is verified with extensive experiments on the unlabeled SurgToolLoc-2022 dataset. Codes and implementation details are available at \url{https://github.com/bhattarailab/Surgical-Vision-World-Model}.

% Authors must provide keywords and are not allowed to remove this Keyword section.

\end{abstract}

\section{Introduction}

The application of artificial intelligence (AI) in surgery has the potential to revolutionize patient care by providing real-time surgical help, simulated training, and tools that support decision making. Recent advancements have already facilitated applications such as object detection \cite{bamba2021automated,yue2024surgicalsam}, and automated critical view of safety assessment \cite{li2023automated,petracchi2024use}. 
Another emerging avenue is the development of autonomous robotic surgical agents by training in simulated environments \cite{tagliabue2020unityflexml,scheikl2022sim}. However, traditional simulations generally fail to account for the complexities of real-world environments and can lead to poor transferability of learned agents \cite{tobin2017domain,kadian2020sim2real}, necessitating the development of realistic interactive environments.
The capability to realistically simulate future states given the current state and action also has other tremendous potential applications, such as providing a risk-free immersive environments to train medical professionals, pre-surgical planning to anticipate complications and test different strategies for optimal outcome, and providing highly personalized approach to surgery, addressing individual needs and mitigating potential risks \cite{sanchez2021application}.
One promising avenue towards this goal is the development of a Surgical World Model, leveraging generative AI to model complex surgical environments and learn to simulate future states based on the patient's current state and surgical actions.

World models build an interactive simulation by modeling the dynamics of the environment, utilizing (state, action, future state) triplets. World models have been extensively studied in the natural domain for developing interactive environments to train reinforcement learning agents \cite{ha2018world,hafner2023mastering} and generating realistic simulations \cite{bruce2024genie}. Recently, a few works have been explored in medical imaging. For instance, Jiang et al. \cite{jiang2024cardiac,jiang2024structure} proposed the use of cardiac world models for the task of probe guidance.  However, the application of the world model to surgical scenarios has been largely unexplored. A recent work by Lin et al. \cite{gas} proposed a world model-based reinforcement learning controller agent for surgical grasping using Dreamer-V2 \cite{dreamerv2} based world model.
However, many of these approaches rely on ground truth action information, the positional difference of tools between each time step, to be accompanied by the state data to train the world model. 
While action data can be obtained from computer simulations, these environments lack the realism needed for effective real-world training. Additionally, using robotic systems to track positions and infer actions is not a scalable solution, as such devices are extremely expensive and not widely available.
On the other hand, labeling real-world surgical videos is also infeasible, given the large amount of data to be labeled, the level of expertise required, and the level of fine-grained annotations required for the multiple surgical tools used \cite{ward2021challenges}. This underscores the requirement of building robust world models utilizing real-world surgical videos without relying on ground truth action data. 

Toward these goals, we draw inspiration from the foundation world model like Genie \cite{bruce2024genie}, which is capable of generating interactive environments from unstructured video data without action annotations. Genie’s latent action model and autoregressive dynamics models enable it to predict future states and infer latent actions purely from visual inputs, making it a promising framework for surgical applications where obtaining action annotations is prohibitively expensive. 
We note that while a surgical world model may be deemed similar to traditional surgical video generation models \cite{cho2024surgen,li2024endora}, such approaches do not facilitate step-wise action conditioning, and are not directly comparable to this work. To the best of our knowledge, this is the first study to explore a foundation world model without action annotation for surgical application. We summarize our key contributions below:
\begin{itemize}
    \item We introduce \ModelName{}, the first action-controllable surgical visual world model. % under resource resource-constraint setting. 
    \item Our experiment on the surgical dataset SurgToolLoc-2022 \cite{surgtoolloc} indicates high-quality generation and controllability, qualitatively and quantitatively.
\end{itemize}

\section{Methodology}
\begin{figure}[htb]
  \centering
  \begin{subfigure}[b]{0.9\linewidth}
    \centering
    \includegraphics[width=\linewidth]{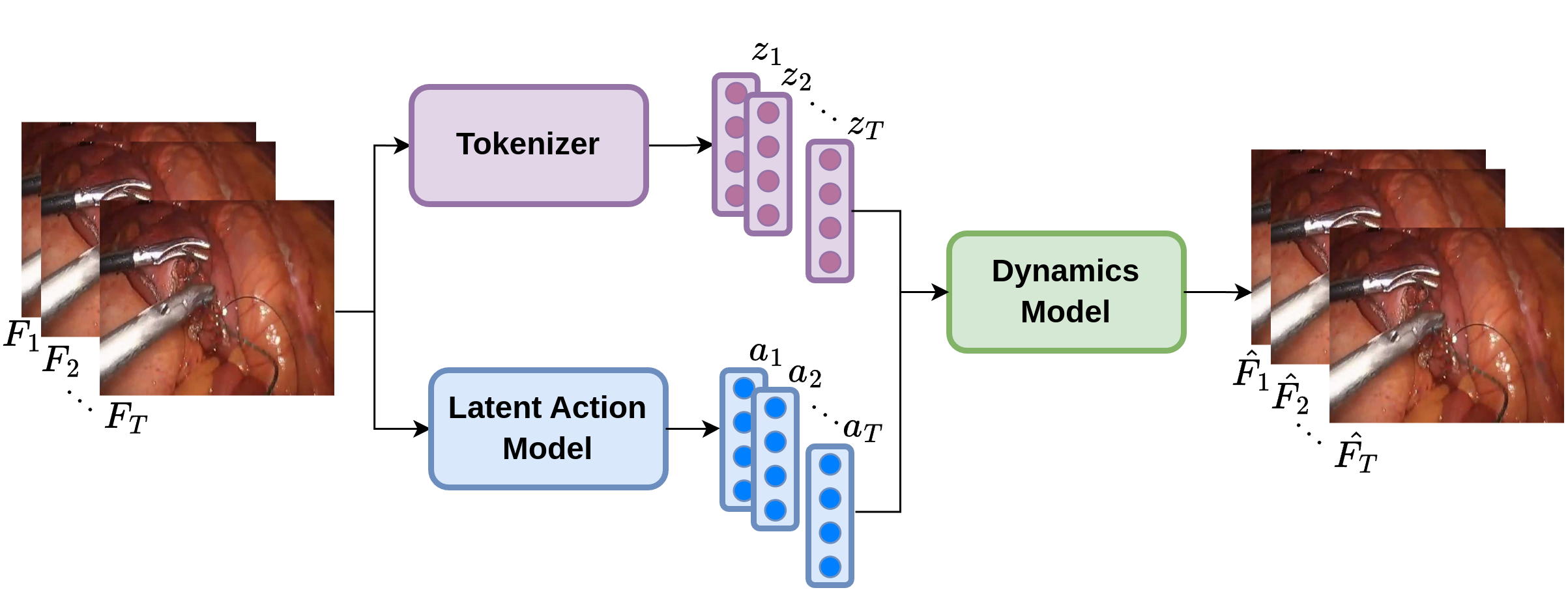}
    \caption{Different Components of \ModelName}
    \label{fig:\ModelName{}}
  \end{subfigure}
  \hfill
  
  \vspace{1em}
  
  \begin{subfigure}[b]{0.48\linewidth}
    \centering
    \includegraphics[width=\linewidth]{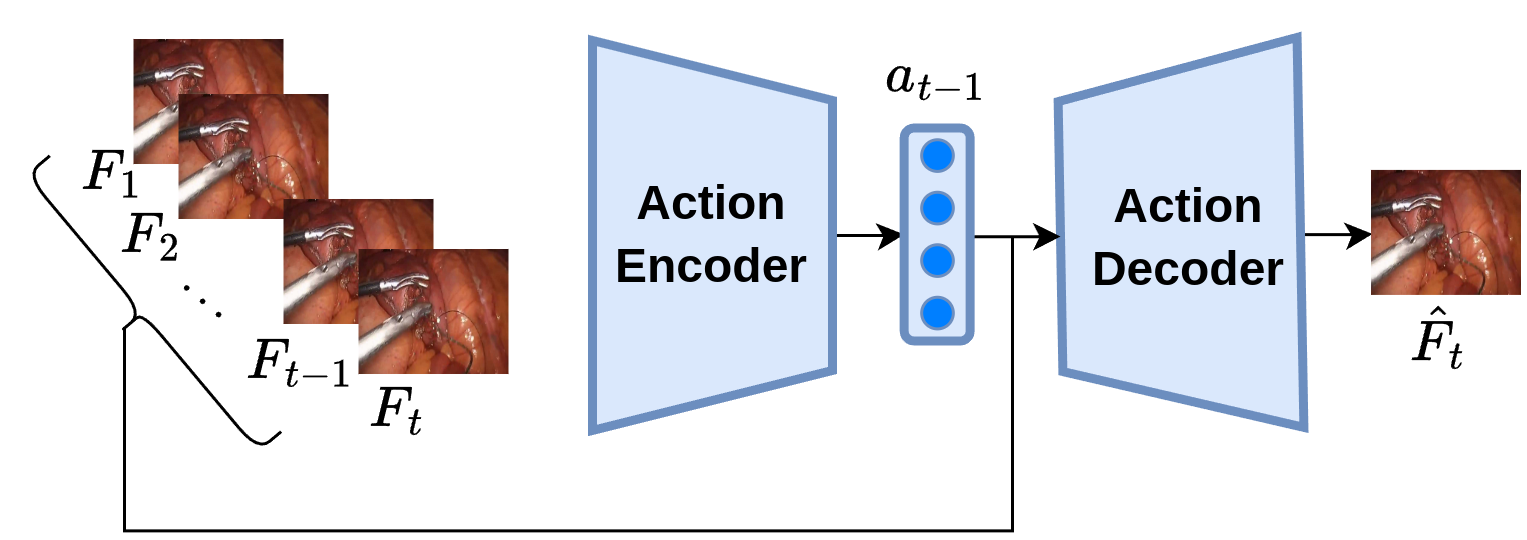}
    \caption{Surgical Latent Action Model}
    \label{fig:lam}
  \end{subfigure}
  \hfill
  \begin{subfigure}[b]{0.42\linewidth}
    \centering
    \includegraphics[width=\linewidth]{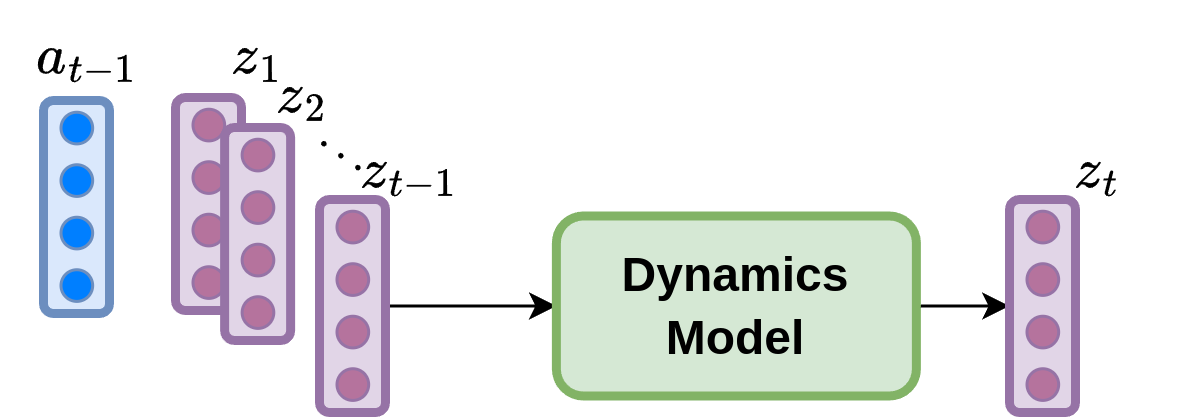}
    \caption{Surgical Dynamics Model}
    \label{fig:dm}
  \end{subfigure}
  
  \caption{Overview of \ModelName{} components and associated models.}
  \label{fig:combined}
\end{figure}
\ModelName{} consists of three key components: the Video Tokenizer, the Surgical Latent Action Model, and the Surgical Dynamics Model (Figure \ref{fig:\ModelName{}}). Each component utilize the spatio-temporal (ST) transformer architecture \cite{st-transformer} that efficiently captures spatial and temporal dependencies, by stacking spatial-only and temporal-only attention within each block. Moreover, its causal temporal attention mechanism facilitates autoregressive training required for future prediction.

\subsubsection{Video Tokenizer}
The video tokenizer encodes 
a sequence of image frames into discrete tokens by leveraging the causal processing of ST-Transformer. It is trained using the standard VQ-VAE objective \cite{van2017neural}. Concretely, given input frames $F_{1:T}$, the ST-Transformer-based encoder causally processes the inputs and produces features $h_{1:T}^v$. The embeddings are quantized to $z_{1:T}^v$, then decoded causally by an ST-Transformer-based decoder to reconstruct the image space $\hat{F}_{1:T}$. The model is trained using a reconstruction objective between the original and predicted images and commitment loss for the encoder. We omit the codebook alignment loss and instead opted for momentum update of codebook vectors as proposed in \cite{van2017neural}. 
The final objective for training the video tokenizer is expressed in Equation \ref{tokenizer_loss}, where $\beta$ is the commitment weight and \textit{sg} refers to the stop-gradient operator: 

\begin{equation}
\mathcal{L}_v(F, \hat{F}) = \|F-\hat{F}\|_2^2 + \beta \|\textit{sg}(z^v)-h^v\|_2^2
\label{tokenizer_loss}
\end{equation}

\subsubsection{Surgical Latent Action Model}
The Latent Action Model (Figure \ref{fig:combined}b) learns to extract latent surgical action features in an unsupervised manner from the video frames. Given input frames $F_{1:T}$, an ST-Transformer based encoder produces features $h_{1:T}^a$. The features $h_{2:T}^a$ are quantized to obtain predicted action embeddings $a_{1:T-1}$. Here, the prediction of $a_i$ is conditioned on the frames $F_{1: i+1}$, and corresponds to the action taken after $F_{1:i}$ to obtain the frame $F_{i+1}$.
The decoder takes the input frames $F_{1:T-1}$ and actions $a_{1:T-1}$ to produce the predictions $\hat{F}_{2:T}$. The surgical latent action model is also trained using the VQ-VAE based objective using reconstruction loss between $F_{2:T}$ and $\hat{F}_{2:T}$, and commitment loss for the encoder, in a similar manner to the video tokenizer. This model only exists to learn latent actions from the data and is not required during inference. The final training objective for the latent action model is: 

\begin{equation}
\mathcal{L}_a(F, \hat{F}) = \|F-\hat{F}\|_2^2 + \beta \|\textit{sg}(a)-h^a\|_2^2
\label{tokenizer_loss}
\end{equation}

\subsubsection{Surgical Dynamics Model}
This component (Figure \ref{fig:combined}c) learns to capture the surgical environment dynamics and is trained to predict the future surgical state, given the present one and the current surgical action embedding. The state information is represented by the tokenized space of the Video Tokenizer. Specifically, the Dynamics Model is an ST-Transformer-based causal transformer that takes in the tokenized video $z_{1:T-1}^v$ from the Video Tokenizer encoder and latent action embeddings $a_{1:T-1}$ from the Latent Action Model as inputs to produce the predictions for future video tokens $z_{2:T}^v$ using masked token prediction objective based on MaskGIT \cite{maskgit}. The resulting tokens are de-tokenized using the Video tokenizer's decoder. During inference, actions can be randomly sampled from the latent action model's codebook as input to the dynamics model. Inference makes use of iterative decoding as proposed in MaskGIT \cite{maskgit}.

\subsection{Training}

The training of the entire generative pipeline consists of two stages. The Video Tokenizer and the Surgical Latent Action Model can be trained simultaneously. The Video Tokenizer is trained to encode a set of input frames to low-dimensional representation. Similarly, the Surgical Latent Action model is trained to extract action information from inputs of the current state and the next. For computational efficiency, we train the latent action model on a lower resolution of 60 x 40 pixels. In the second stage, the Surgical Dynamics Model is trained to predict tokenized features of the future state given the past frames and action vectors predicted by the latent action model.

\section{Implementation Details}
\subsubsection{Data}
We utilize the SurgToolLoc-2022 dataset \cite{surgtoolloc} for training \ModelName{}. The dataset consists of video clips taken from surgical training exercises using the da Vinci robotic system, and showcase surgical trainees performing standard activities such as dissecting tissue and suturing. The dataset consists of 30-seconds long 24,695 video clips, captured at 60 fps at a resolution of 1280 x 720, from one channel of the endoscope. For the extent of each clip, three robotic surgical tools out of 14 possible ones are installed, and within the surgical field. 

All models are trained on 16 frame clips sampled at 1 fps from the videos. A center crop of 900 x 600 pixels is applied to remove black borders and possible digital overlay. The action model is trained by resizing the image to 60 x 40 pixels and the tokenizer is trained by resizing the image to 120 x 180 pixels. Training the action model at a lower resolution helps to significantly decrease the compute requirement.
Additionally, our early experiments suggested that training the action model at a lower resolution did not degrade the quality of samples generated by the dynamics model at a higher resolution. We refer to Tables \ref{tab:combined} for the different hyperparameters used in training \ModelName{}.

\begin{table}[htb]
  \centering
  \caption{Hyperparameters for different components} 
  
  \begin{subtable}[b]{0.45\linewidth}
    \centering
    \caption{Hyperparameters for Video Tokenizer}
    
    \renewcommand{\arraystretch}{1} % Adjust row spacing
    \setlength{\tabcolsep}{4pt}
    \begin{tabular}{l l c}
      \toprule
      \textbf{Component} & \textbf{Parameter} & \textbf{Value} \\
      \midrule
      \multirow{3}{*}{Encoder} & num\_layers & \textbf{4} \\
                             & d\_model    & \textbf{384} \\
                             & num\_heads  & \textbf{12} \\
      \midrule
      \multirow{3}{*}{Decoder} & num\_layers & \textbf{6} \\
                             & d\_model    & \textbf{384} \\
                             & num\_heads  & \textbf{12} \\
      \midrule
      \multirow{3}{*}{Codebook} & num\_codes  & \textbf{1024} \\
                              & patch\_size & \textbf{(4, 4)} \\
                              & latent\_dim & \textbf{32} \\
      \midrule
      \multirow{3}{*}{Training} & learning\_rate  & $\mathbf{10^{-4}}$ \\
                              & $\beta_1$ & \textbf{0.9} \\
                              & $\beta_2$ & \textbf{0.9999} \\
      \bottomrule
    \end{tabular}
    \label{tab:sthyperparameter}
  \end{subtable}
  \hfill
  \begin{subtable}[b]{0.45\linewidth}
    \centering
    \renewcommand{\arraystretch}{1} % Adjust row spacing
    \setlength{\tabcolsep}{4pt}
    \caption{Hyperparameters for Surgical Latent Action Model}
    
    \begin{tabular}{l l c}
      \toprule
      \textbf{Component} & \textbf{Parameter} & \textbf{Value} \\
      \midrule
      \multirow{3}{*}{Encoder} & num\_layers & \textbf{8} \\
                             & d\_model    & \textbf{384} \\
                             & num\_heads  & \textbf{12} \\
      \midrule
      \multirow{3}{*}{Decoder} & num\_layers & \textbf{12} \\
                             & d\_model    & \textbf{384} \\
                             & num\_heads  & \textbf{12} \\
      \midrule
      \multirow{3}{*}{Codebook} & num\_codes  & \textbf{12} \\
                              & patch\_size & \textbf{(4, 4)} \\
                              & latent\_dim & \textbf{32} \\
      \midrule
      \multirow{3}{*}{Training} & learning\_rate  & $\mathbf{10^{-5}}$ \\
                              & $\beta_1$ & \textbf{0.9} \\
                              & $\beta_2$ & \textbf{0.9999} \\
      \bottomrule
    \end{tabular}
    \label{tab:lahyperparameter}
  \end{subtable}
  
  \vspace{1em}
  
  \begin{subtable}[b]{1.0\linewidth}
    \caption{Hyperparameters for Surgical Dynamics Model}
  
    \centering
    \renewcommand{\arraystretch}{1} % Adjust row spacing
    \setlength{\tabcolsep}{4pt}
    \begin{tabular}{l c c c c c}
      \toprule
      \textbf{Parameters} & \textbf{num\_layers} & \textbf{num\_heads} & \textbf{d\_model} & \textbf{FLOPs} \\
      \midrule
      62.5M & 12 & 8 & 512 & \(14 \times 10^{18}\) \\
      \bottomrule
    \end{tabular}
    \label{tab:dmhyperparameter}
  \end{subtable}
  
  \label{tab:combined}
\end{table}

\section{Results}

\subsection{Qualitative Results}
In this section, we present sample generations produced by \ModelName{}. The model can generate new frames given one or more prompt frames and action embedding. The action embeddings for generating frame $F_{t+1}$ can be obtained from the Surgical Latent Action model by either randomly sampling the codebook vectors (non ground-truth trajectory), or inferred from the ground truth (GT) frame $F_{t+1}$ using the latent action model.
All results were produced by sampling with 1.0 temperature and 25 maskgit steps.
\begin{figure}[H]
    \centering
    \includegraphics[width=\linewidth]{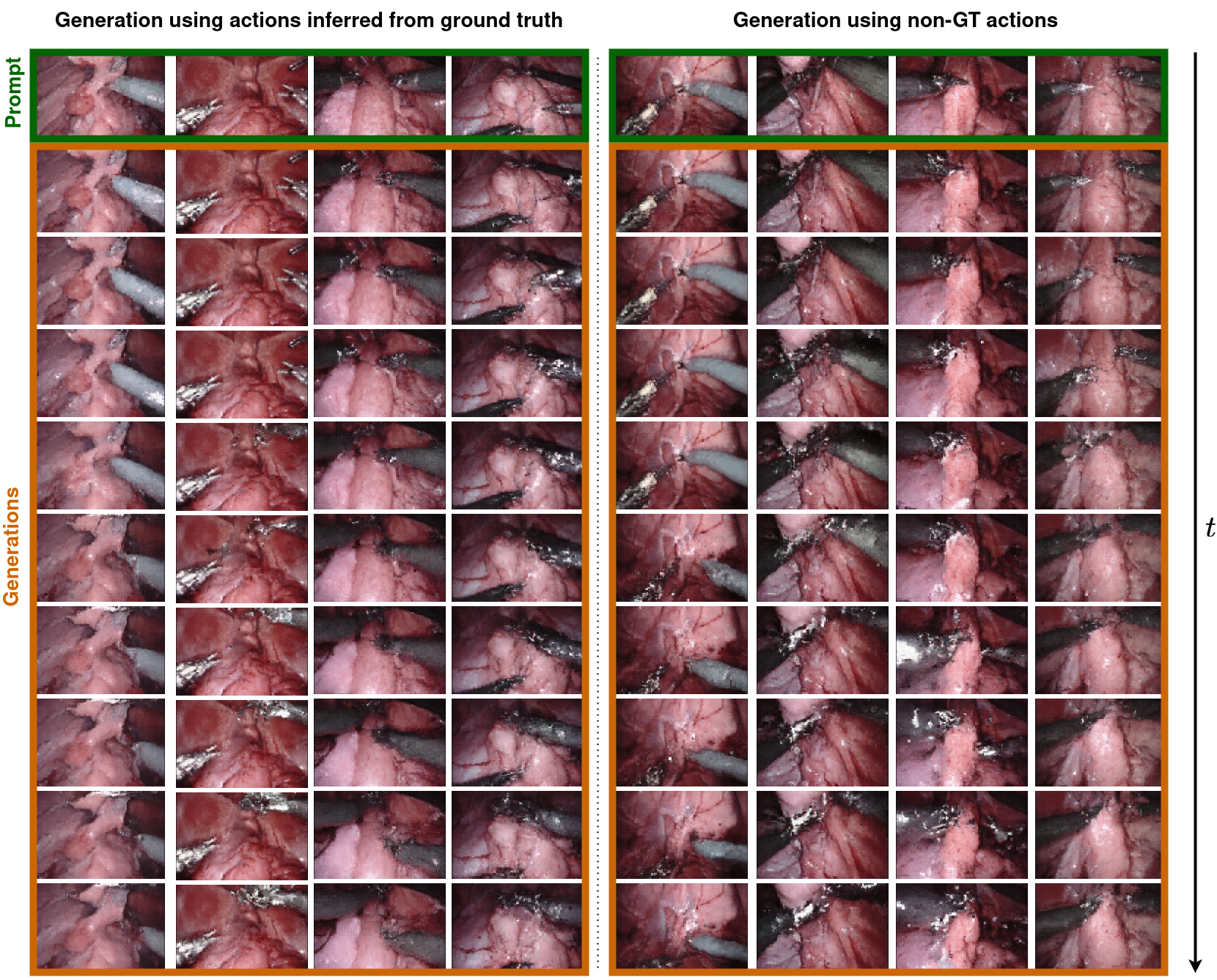}
    \caption{Generation using a single frame prompt, based on actions inferred from ground truth(left) and non ground truth trajectory by random sampling (right). Frames shown are generated at 1fps interval.}
    \label{fig:gt_rand}
\end{figure}
In Figure \ref{fig:gt_rand}, we present generation samples obtained from \ModelName{}, across a variety of prompt frames. All generations are obtained from a single starting prompt frame and a sequence of action embeddings autoregressively. The images on the left present generations using actions that follow ground truth trajectory, while the images on the right are generated by sampling actions to follow different trajectory. As observed, all generations preserve the original surgical field accurately and show movement in surgical tools, while mostly preserving their shape across frames. Notably, the model is also capable of identifying and realistically modeling reflective tools and their interactions with the tissues. We can also observe tissue deformation in response to the tool actions in the first column of non-GT generations in figure \ref{fig:gt_rand}.

\begin{figure}[H]
    \centering
    \includegraphics[width=\linewidth]{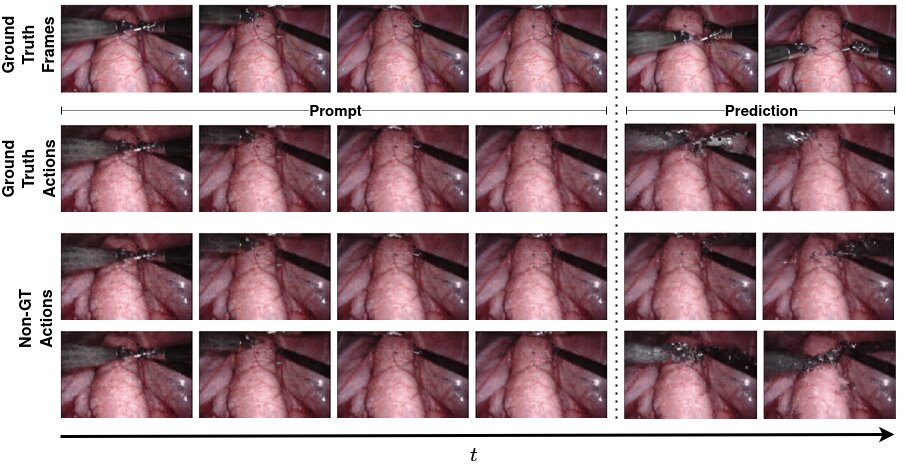}
    \caption{Generation using actions inferred from ground truth vs. non ground truth actions}
    \label{fig:action}
\end{figure}

Figure \ref{fig:action} shows an example of prediction of two new frames given four prompt frames.  As observed, \ModelName{}  is capable of generating frames maintaining consistency in the surgical environment, shape of tools, and capturing movement between frames. As shown in the figure, some of the tools are only visible in the first couple of frames in the prompt, but the model predicts reappearance of both tools into the frame using the action inferred from the ground truth by the latent action model. We can also see alternate trajectories of the tools produced by conditioning on different actions. 

We additionally present 
some generation samples in the form of a video in the github repository\footnote{\href{https://github.com/bhattarailab/Surgical-Vision-World-Model/blob/main/examples}{https://github.com/bhattarailab/Surgical-Vision-World-Model/blob/main/examples}}. We notice that the model is also capable of accurately capturing the natural pulsing behavior of the surgical field, caused by respiratory motion of the body. 
Thus, the generations obtained by our model, \ModelName{} are able to capture different aspects of real world surgical data. Additionally, generations vary when conditioned on different latent action embeddings, highlighting action-conditioned generation capability of \ModelName{}, which is crucial to enable applications like robust training of autonomous surgical agents and realistic training for medical professionals.

\subsection{Quantitative Results}

We examine the performance of our model on two criteria, quality of frame generation and controllability. 
We use Fréchet Video Distance \textbf{(FVD)} and  Structural Similarity Index \textbf{(SSIM)}  to measure generation quality. FVD is a video-level metric and has high alignment with human evaluation \cite{unterthiner2019fvd}. SSIM considers luminance, contrast, and structural similarities, making it effective for assessing how viewers perceive quality \cite{wang2004image}. To measure the controllability aspect of \ModelName{}, we use \textbf{$\Delta$PSNR}, following Genie \cite{bruce2024genie}. This metric measures the extent to which generated videos differ when conditioned on latent actions inferred from ground truth vs. when sampled from a random action distribution. 
The FVD was calculated on a total of 10 frames, including the prompt frames.

In Table \ref{tab:quantresults}, we present two sets of results: when generation is conditioned based on a single prompt frame and when conditioned on 4 prompt frames. We observe that conditioning the generation of ground truth action compared to a random distribution results in a positive $\Delta$PSNR value. This shows a definite difference in generation based on conditioning, highlighting the controllability aspect of \ModelName{}. Furthermore, we observe better SSIM and FVD values when conditioned on actions inferred from ground truth, suggesting that our surgical latent action models is able to capture the latent action information. We also observe better generation quality, in terms of $\Delta$PSNR and SSIM scores when conditioned on additional prompt frames.

\begin{table}[H]
\caption{Quantitative Results.}
\centering
\renewcommand{\arraystretch}{1.2} % Adjust row spacing
\setlength{\tabcolsep}{4pt}
\begin{tabular}{cccc|ccc|c}
\hline
& \multicolumn{3}{c|}{\textbf{PSNR ($\uparrow$)}} & \multicolumn{3}{c|}{\textbf{SSIM ($\uparrow$)}} & $\mathbf{FVD_{10}}$ ($\downarrow$) \\ \hline
No. of frames generated & 2 & 4 & 6 & 2 & 4 & 6 & - \\ \toprule
\multicolumn{8}{l}{{\textbf{Prompt Frames: 1}}} \\ \hline
GT action & \textbf{17.67} & \textbf{17.05} & \textbf{16.49} & \textbf{0.44} & \textbf{0.42} & \textbf{0.39} & \textbf{1717.59} \\
Non-GT action & 15.86	& 15.10 & 14.73 & 0.37 & 0.33 & 0.30 & 2079.46 \\ \hline
$\Delta$PSNR$(\uparrow)$ & 1.81 & 1.95 & 1.76 & - & - & - & - \\ \toprule
\multicolumn{8}{l}{{\textbf{Prompt Frames: 4}}} \\ \hline
GT action & \textbf{18.74} & \textbf{17.82} & \textbf{17.23} & \textbf{0.52} & \textbf{0.49} & \textbf{0.45} & \textbf{1290.17} \\
Non-GT action & 16.67 & 15.75 & 15.15 & 0.44 & 0.39 & 0.35 & 1382.74 \\ \hline
$\Delta$PSNR$(\uparrow)$ & 2.07 & 2.07 & 2.08 & - & - & - & - \\ \hline
\end{tabular}
\label{tab:quantresults}
\end{table}

\section{Conclusion}
In this work, we presented the first study building a surgical world model utilizing raw surgical videos without any action data. We obtain high quality generation ability from \ModelName{}, producing prompt consistent surgical frames with movement in tools across frames. 
Additionally, we highlight the model's ability to condition generation during inference, based on action embeddings at each time sequence. Future work could explore training an RL agent in generated environments, refining the latent action model to better disentangle actions, improve tool shape consistency, and explore semi-supervised approaches to learning the latent action model-- leveraging small available action annotated datasets.

% \newpage

\bibliographystyle{splncs04}
\bibliography{references}

\begin{thebibliography}{10}
\providecommand{\url}[1]{\texttt{#1}}
\providecommand{\urlprefix}{URL }
\providecommand{\doi}[1]{https://doi.org/#1}

\bibitem{bamba2021automated}
Bamba, Y., Ogawa, S., Itabashi, M., Kameoka, S., Okamoto, T., Yamamoto, M.: Automated recognition of objects and types of forceps in surgical images using deep learning. Scientific Reports  \textbf{11}(1),  22571 (2021)

\bibitem{bruce2024genie}
Bruce, J., Dennis, M.D., Edwards, A., Parker-Holder, J., Shi, Y., Hughes, E., Lai, M., Mavalankar, A., Steigerwald, R., Apps, C., et~al.: Genie: Generative interactive environments. In: Forty-first International Conference on Machine Learning (2024)

\bibitem{maskgit}
Chang, H., Zhang, H., Jiang, L., Liu, C., Freeman, W.T.: Maskgit: Masked generative image transformer (2022), \url{https://arxiv.org/abs/2202.04200}

\bibitem{cho2024surgen}
Cho, J., Schmidgall, S., Zakka, C., Mathur, M., Kaur, D., Shad, R., Hiesinger, W.: Surgen: Text-guided diffusion model for surgical video generation. arXiv preprint arXiv:2408.14028  (2024)

\bibitem{ha2018world}
Ha, D., Schmidhuber, J.: World models. arXiv preprint arXiv:1803.10122  (2018)

\bibitem{dreamerv2}
Hafner, D., Lillicrap, T., Norouzi, M., Ba, J.: Mastering atari with discrete world models (2022), \url{https://arxiv.org/abs/2010.02193}

\bibitem{hafner2023mastering}
Hafner, D., Pasukonis, J., Ba, J., Lillicrap, T.: Mastering diverse domains through world models. arXiv preprint arXiv:2301.04104  (2023)

\bibitem{jiang2024structure}
Jiang, H., Li, M., Sun, Z., Jia, N., Sun, Y., Luo, S., Song, S., Huang, G.: Structure-aware world model for probe guidance via large-scale self-supervised pre-train. In: International Workshop on Advances in Simplifying Medical Ultrasound. pp. 58--67. Springer (2024)

\bibitem{jiang2024cardiac}
Jiang, H., Sun, Z., Jia, N., Li, M., Sun, Y., Luo, S., Song, S., Huang, G.: Cardiac copilot: Automatic probe guidance for echocardiography with world model. In: International Conference on Medical Image Computing and Computer-Assisted Intervention. pp. 190--199. Springer (2024)

\bibitem{kadian2020sim2real}
Kadian, A., Truong, J., Gokaslan, A., Clegg, A., Wijmans, E., Lee, S., Savva, M., Chernova, S., Batra, D.: Sim2real predictivity: Does evaluation in simulation predict real-world performance? IEEE Robotics and Automation Letters  \textbf{5}(4),  6670--6677 (2020)

\bibitem{li2024endora}
Li, C., Liu, H., Liu, Y., Feng, B.Y., Li, W., Liu, X., Chen, Z., Shao, J., Yuan, Y.: Endora: Video generation models as endoscopy simulators. In: International Conference on Medical Image Computing and Computer-Assisted Intervention. pp. 230--240. Springer (2024)

\bibitem{li2023automated}
Li, Y., Gupta, H., Ling, H., Ramakrishnan, I., Prasanna, P., Georgakis, G., Sasson, A.: Automated assessment of critical view of safety in laparoscopic cholecystectomy. In: 2023 IEEE 11th International Conference on Healthcare Informatics (ICHI). pp. 330--337. IEEE (2023)

\bibitem{gas}
Lin, H., Li, B., Wong, C.W., Rojas, J., Chu, X., Au, K.W.S.: World models for general surgical grasping (2024), \url{https://arxiv.org/abs/2405.17940}

\bibitem{petracchi2024use}
Petracchi, E.J., Olivieri, S.E., Varela, J., Canullan, C.M., Zandalazini, H., Ocampo, C., Quesada, B.M.: Use of artificial intelligence in the detection of the critical view of safety during laparoscopic cholecystectomy. Journal of Gastrointestinal Surgery  \textbf{28}(6),  877--879 (2024)

\bibitem{sanchez2021application}
S{\'a}nchez-Margallo, J.A., Plaza~de Miguel, C., Fern{\'a}ndez~Anzules, R.A., S{\'a}nchez-Margallo, F.M.: Application of mixed reality in medical training and surgical planning focused on minimally invasive surgery. Frontiers in Virtual Reality  \textbf{2},  692641 (2021)

\bibitem{scheikl2022sim}
Scheikl, P.M., Tagliabue, E., Gyenes, B., Wagner, M., Dall'Alba, D., Fiorini, P., Mathis-Ullrich, F.: Sim-to-real transfer for visual reinforcement learning of deformable object manipulation for robot-assisted surgery. IEEE Robotics and Automation Letters  \textbf{8}(2),  560--567 (2022)

\bibitem{tagliabue2020unityflexml}
Tagliabue, E., Pore, A., Dall’Alba, D., Piccinelli, M., Fiorini, P., et~al.: Unityflexml: Training reinforcement learning agents in a simulated surgical environment. In: I-RIM 2020 conference proceedings. pp.~0--1 (2020)

\bibitem{tobin2017domain}
Tobin, J., Fong, R., Ray, A., Schneider, J., Zaremba, W., Abbeel, P.: Domain randomization for transferring deep neural networks from simulation to the real world. In: 2017 IEEE/RSJ international conference on intelligent robots and systems (IROS). pp. 23--30. IEEE (2017)

\bibitem{unterthiner2019fvd}
Unterthiner, T., Van~Steenkiste, S., Kurach, K., Marinier, R., Michalski, M., Gelly, S.: Fvd: A new metric for video generation  (2019)

\bibitem{van2017neural}
Van Den~Oord, A., Vinyals, O., et~al.: Neural discrete representation learning. Advances in neural information processing systems  \textbf{30} (2017)

\bibitem{wang2004image}
Wang, Z., Bovik, A.C., Sheikh, H.R., Simoncelli, E.P.: Image quality assessment: from error visibility to structural similarity. IEEE transactions on image processing  \textbf{13}(4),  600--612 (2004)

\bibitem{ward2021challenges}
Ward, T.M., Fer, D.M., Ban, Y., Rosman, G., Meireles, O.R., Hashimoto, D.A.: Challenges in surgical video annotation. Computer Assisted Surgery  \textbf{26}(1),  58--68 (2021)

\bibitem{st-transformer}
Xu, M., Dai, W., Liu, C., Gao, X., Lin, W., Qi, G.J., Xiong, H.: Spatial-temporal transformer networks for traffic flow forecasting (2021), \url{https://arxiv.org/abs/2001.02908}

\bibitem{yue2024surgicalsam}
Yue, W., Zhang, J., Hu, K., Xia, Y., Luo, J., Wang, Z.: Surgicalsam: Efficient class promptable surgical instrument segmentation. In: Proceedings of the AAAI Conference on Artificial Intelligence. vol.~38, pp. 6890--6898 (2024)

\bibitem{surgtoolloc}
Zia, A., Bhattacharyya, K., Liu, X., Berniker, M., Wang, Z., Nespolo, R., Kondo, S., Kasai, S., Hirasawa, K., Liu, B., Austin, D., Wang, Y., Futrega, M., Puget, J.F., Li, Z., Sato, Y., Fujii, R., Hachiuma, R., Masuda, M., Saito, H., Wang, A., Xu, M., Islam, M., Bai, L., Pang, W., Ren, H., Nwoye, C., Sestini, L., Padoy, N., Nielsen, M., Schüttler, S., Sentker, T., Husseini, H., Baltruschat, I., Schmitz, R., Werner, R., Matsun, A., Farooq, M., Saaed, N., Viera, J.R.R., Yaqub, M., Getty, N., Xia, F., Zhao, Z., Duan, X., Yao, X., Lou, A., Yang, H., Han, J., Noble, J., Wu, J.Y., Alshirbaji, T.A., Jalal, N.A., Arabian, H., Ding, N., Moeller, K., Chen, W., He, Q., Bilal, M., Akinosho, T., Qayyum, A., Caputo, M., Vohra, H., Loizou, M., Ajayi, A., Berrou, I., Niyi-Odumosu, F., Maier-Hein, L., Stoyanov, D., Speidel, S., Jarc, A.: Surgical tool classification and localization: results and methods from the miccai 2022 surgtoolloc challenge (2023), \url{https://arxiv.org/abs/2305.07152}

\end{thebibliography}

\end{document}